\documentclass[12pt]{article}
\usepackage{amsmath, amsfonts, amssymb, amsxtra, amsfonts, amsthm, bbold, subcaption, xcolor, enumerate, natbib}
\usepackage{graphicx}
\usepackage[margin = 1in]{geometry}

\usepackage{enumerate}
\usepackage{xcolor} 
\newtheorem{assumption}{Assumption}
\usepackage{multirow}
\usepackage{threeparttable}
\usepackage{booktabs}
\usepackage{amsmath, amssymb}
\usepackage{tikz}
\usetikzlibrary{positioning,shapes.geometric,external}

\definecolor{dacolor}{RGB}{0,0,0} 
\definecolor{highlightcolor}{RGB}{0,0,0}

\newcommand{\pa}{\mathrm{pa}}
\newcommand{\piyi}{\pi_{Y, i}}
\newcommand{\pimi}{\pi_{M, i}}
\newcommand{\piui}{\pi_{U, i}}

\newcommand{\calL}{\mathcal{L}}

\newcommand{\cindep}{\mathrel{\text{\scalebox{1.07}{$\perp\mkern-10mu\perp$}}}}

\newcommand{\logit}[1]{\mathrm{logit}\left( #1 \right)}

\newcommand{\E}[1]{\mathbb{E}\left[ #1 \right]}
\newcommand{\logitinv}[1]{\mathrm{logit}^{-1}\left( #1 \right)}

\usepackage[figuresright]{rotating}

\title{Bayesian data fusion for unmeasured confounding}

\author{Leah Comment, Brent A. Coull, Corwin Zigler, Linda Valeri}

\begin{document}


\date{\today}



\label{firstpage}

\maketitle 

\begin{abstract}
Bayesian causal inference offers a principled approach to policy evaluation of proposed interventions on mediators or time-varying exposures. We outline a general approach to the estimation of causal quantities for settings with time-varying confounding, such as exposure-induced mediator-outcome confounders. We further extend this approach to propose two Bayesian data fusion (BDF) methods for unmeasured confounding. Using informative priors on quantities relating to the confounding bias parameters, our methods incorporate data from an external source where the confounder is measured in order to make inferences about causal estimands in the main study population. We present results from a simulation study comparing our data fusion methods to two common frequentist correction methods for unmeasured confounding bias in the mediation setting. We also demonstrate our method with an investigation of the role of stage at cancer diagnosis in contributing to Black-White colorectal cancer survival disparities.
\end{abstract}

\newbox\keywdbox
\def\keywords{\global\setbox\keywdbox\vbox\bgroup\hsize\textwidth\small\leftskip0pc\rightskip\leftskip\noindent{\sc
Key words:\hskip1em}\ignorespaces}
\def\endkeywords{\egroup}

\begin{keywords}
Causal inference; Data fusion; g-formula; Mediation; Racial disparities; Unmeasured confounding.
\end{keywords}


\section{Introduction}
The value of causal evidence from a statistical analysis depends on the quality and suitability of the data source. In the era of Big Data, many large data sources valuable for health research are used in ways that were not foreseen by the original collectors. As such, these sources are missing one or more key covariates. For example, electronic health records and tumor registries may lack important socioeconomic and behavioral factors. If these unmeasured factors act as confounders of the relationship(s) of interest, causal quantities may not be estimated using the observed data, regardless of the sample size. 

Widespread availability of large samples has also ushered in more sophisticated statistical models for disentangling causal effects. Even when many covariates are measured, analyses that take advantage of the rich, longitudinal nature of data sources like electronic health records are vulnerable to time-varying confounding by unmeasured variables. Proper control of confounding is particularly difficult when exposure status changes over time, with later exposure determined in part by covariates influenced by previous exposure. This phenomenon appears in almost every medical setting as doctors tailor treatment based on patient history and current health state. Analogous problems arise in the context of mediation analysis, within which the temporal ordering of treatments, mediators, and outcomes can yield structures analogous to time-varying exposures.

When important confounders are unavailable, researchers typically conduct sensitivity analyses to assess whether bias due to the unmeasured confounding is likely to alter the substantive conclusions of the research. Recent methodological advances have identified sharp nonparametric bounds for common causal estimands such as the average treatment effect \citep{ding2016sensitivity} as well as various mediation quantities \citep{ding2016sharp}. Several bias correction formulae provide adjusted point estimates and confidence intervals based on bias values found in the literature \citep{vanderweele2014sensitivity, vanderweele2015explanation}. In the absence of information about the sources of confounding, they can be used to determine the strength of confounding needed to eliminate statistical significance. Individual approaches also require rare outcomes, specific link functions, or assumptions about effect modification \citep{vanderweele2015explanation}. With some notable exceptions \citep{mccandless2017bayesian, greenland2005multiple}, correction methods rarely incorporate uncertainty surrounding the bias parameters. Generally speaking, most existing sensitivity methods suffer from poor extensibility to both arbitrary confounding structures and longitudinal settings.

Fortunately, the era of Big Data is also the era of abundant data. Relationships among the outcome, confounders, and exposure of interest can be found in alternative data sets, though these sources may not be as representative of the target population as the main source. A literature on data fusion methods has arisen to meet the need to combine information from multiple sources. Recent authors have proposed Bayesian variable selection methods with validation data sets \citep{antonelli2017guided} and data integration for information from different scales, such as individual-level and community-level data  \citep{jackson2006improving}.

To address the limited extensibility and uncertainty quantification of existing methods, we propose a general framework for incorporating information from external data sources to perform sensitivity analyses for unmeasured confounding. We develop a Bayesian method for data fusion and incorporate it into an existing Bayesian g-formula approach which adjusts for confounding using parametric models for covariate standardization \citep{keil2015bayesian}. To handle unmeasured confounding, we introduce a procedure for generating informative priors using external data sources. We then describe two estimation strategies to account for unmeasured confounding: one using mechanics similar to \citet{keil2015bayesian}, and the other augmenting this approach with a Bayesian bootstrap procedure for marginalization. We compare such strategies with traditional sensitivity analysis approaches, paying particular attention to potential violations of causal transportability \citep{pearl2011transportability} when the underlying causal processes differ between the two populations from which the main and external data were sampled. Unlike existing approaches, these Bayesian g-formula methods generalize to accommodate unmeasured confounding of many different types, including time-varying confounding found in mediation and analysis of longitudinal treatments.

 Our paper is organized as follows. Section 2 develops a Bayesian g-method that accommodates dynamic and stochastic treatment assignment mechanisms, then highlights the connection between this model and a mediation analysis. We describe two estimators based on the g-formula. Section 3 introduces two Bayesian data fusion algorithms to implement sensitivity analyses for unmeasured confounding. A simulation study comparing the two methods to traditional sensitivity analysis approaches is given in Section 4. In Section 5, we use the data fusion method from Section 3 to augment cancer registry data with information from a cohort study in order to evaluate the role of stage at diagnosis in explaining Black-White disparities in colorectal cancer survival. We conclude with a discussion in Section 6.
 
\section{The Bayesian g-formula for static and dynamic regimes without unmeasured confounding}
\subsection{Causal notation and assumptions}
Let $Y$ denote the observed outcome of interest in a causal graph $G$, with the central scientific question concerning two intervention regimes $g$ and $g'$. One of these regimes may correspond to the ``natural'' assignment mechanism that generated the observed data. Let $V$ be the intervention set (i.e., the nodes intervened upon by either $g$ or $g'$), and let $Z$ be the set of baseline confounders and post-treatment variables \emph{not} influenced by treatment. Let $W$ be the set containing all other nodes in $G$, in which case $W$ includes any variable influenced by treatment but not of primary interest (i.e., not the outcome $Y$) or directly intervened upon (i.e., $W \not\in V$). The complete set of observed data is $O = (Z, V, W, Y)$. Let $Y^g$ denote the potential outcome for $Y$ under regime $g$, with the causal contrast of interest $\tau = \E{Y^g - Y^{g'}}$. \textcolor{highlightcolor}{Depending on the specifics of the regime, mediators of the $A \to Y$ relationship may either be in the set $W$ or $V$. Potential outcomes for $W$ and $V$ under regime $g$ are denoted by $W^g$ and $V^g$. When one mediator is of primary interest, we will denote it by $M$.} Throughout this paper, we must assume Bayesian analogs to positivity (Assumption 1) and consistency (Assumption 2) \citep{keil2015bayesian}. We also require exchangeability conditional on observed variables (Assumption 3) and correct specification of all parametric models (Assumption 4). Formal statements of these assumptions can be found in the supplemental materials.

\subsection{Static and deterministic treatment regimes}
After adopting parametric models indexed by the parameter vector $\theta$, the Bayesian g-formula algorithm outlined by \citet{keil2015bayesian} gives the posterior predictive distribution for a newly observed outcome $Y$ under intervention regime $g_0 \in \{g, g' \}$ as
\begin{equation*}
p(\tilde{y}^{g_0} | o ) = \int p(\tilde{y}^{g_0} | \theta, o) \pi(\theta|o) d\theta
\end{equation*}
where $\pi(\theta|o)$ is the posterior of the parameters $\theta$ given the observed data $O$. 
The posterior distribution of the causal effect $\tau$ is therefore
\begin{equation*}
p(\tau| o) = \int \left( p(\tilde{y}^g | \theta, o) - p(\tilde{y}^{g'} | \theta, o) \right) \pi(\theta|o) d\theta .
\end{equation*}
Keil and colleagues outline a simulation-based algorithm for estimating causal contrasts for static regimes. To facilitate our extension to the mediation setting, we introduce different notation to emphasize the distinction between covariates $Z$ that are unaffected by treatment and covariates $W$ that are caused by one or more variables in the intervention set. For parametric models $p(z | \theta_Z)$, $p(w| v, z, \theta_W)$, and $p(y | w, v, z, \theta_Y)$, the respective parameter likelihoods are $\mathcal{L}(\theta_Z | z)$, $\mathcal{L}(\theta_W | w, v, z)$, and $\mathcal{L}(\theta_Y | y, w, v, z)$. Then the likelihood for the complete parameter vector $\theta = (\theta_Z, \theta_W, \theta_Y)$ is given by
\begin{equation*}
\mathcal{L}(\theta|o) = 
\mathcal{L}(\theta_Y | y, w, v, z)
\times 
\mathcal{L}(\theta_W | w, v, z) 
\times 
\mathcal{L}(\theta_Z | z).
 \label{eqn:staticlike}
\end{equation*}
We assume that $\theta_Y$, $\theta_W$, $\theta_Z$ are independent a priori such that $\pi(\theta) = \pi(\theta_Y) \times \pi(\theta_W) \times \pi(\theta_Z)$. The resulting parameter posterior is $\pi(\theta|o) \propto \mathcal{L}(\theta|o) \times \pi(\theta)$, and the posterior predictive distribution for the outcome under regime $g_0 \in \{g, g'\}$ is
\begin{align}
p(\tilde{y}^{g_0} | o) = \idotsint &
%
p(\tilde{y} | g_0, \tilde{w}, \tilde{z}, \theta_Y)
p(\tilde{w} | g_0, \tilde{z}, \theta_W)
p(\tilde{z} | \theta_Z)
\pi(\theta|o)
d\theta
d\tilde{w}
d\tilde{z}.\label{eqn:bgform}
\end{align}

To distinguish Keil's computationally intensive approach from classical covariate standardization techniques using the g-formula, we refer to this approach as \emph{simulation-based} Bayesian g-formula (BGF-SIM). We now develop a procedure to accommodate dynamic and stochastic treatment regimes.

\subsection{Dynamic and stochastic treatment regimes}
\begin{figure}
		\centering 
		\begin{tikzpicture}[ ->,shorten >=2pt,>=stealth,node distance=1cm,pil/.style={->,thick,shorten =2pt}]
			\node (a) {$A$};
            \node[above left=of a] (z) {$Z$};
			\node[right=of a] (m) {$M$};
			\node[right=of m] (y) {$Y$};
			\node[below=of m] (c) {$W$};
            \draw[->] (z) to (a);
            \draw[->] (z) to (m);
            \draw[->] (z) to [out=10, in=120] (y);
            \draw[->] (z) to [out=270, in=180] (c);
			\draw[->] (a) to [out=30, in=150] (y);
			\draw[->] (a) to (m);
			\draw[->] (a) to (c);
			\draw[->] (m) to (y);
			\draw[->] (c) to (y);
			\draw[->] (c) to (m);
		\end{tikzpicture}
		\caption{Mediation causal structure with outcome $Y$, exposure $A$, mediator $M$, baseline confounder(s) $Z$, and exposure-induced mediator-outcome confounder(s) $W$}
		\label{fig:meddag}
\end{figure}
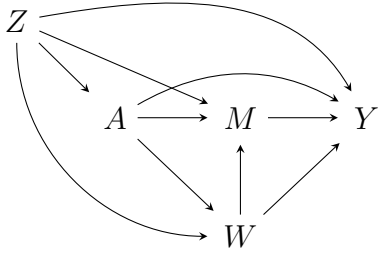

Scientific questions of interest sometimes involve contrasts of regimes which assign exposure stochastically according to different distributions depending on prior covariates. In particular, we may be interested in the ``natural'' assignment mechanism generating the observed data. This is exactly the case for mediation analysis, which decomposes the effect of an exposure on an outcome into component causal pathways in order to understand possible mechanisms enacting the overall effect. Figure~\ref{fig:meddag} shows a classic causal structure in mediation, where the mediator $M$ channels part of the effect of the exposure $A$ on the outcome $Y$, with $Z$ as a baseline confounder. The set $W$ contains other mediators which also act as exposure-induced mediator-outcome confounders for the $M \to Y$ relationship. 

All common mediation estimands can be formulated as contrasts in regimes, including controlled direct effects ($CDE$), natural direct and indirect effects ($NDE$ and $NIE$), and randomized interventional analogs to these quantities \citep{didelez2012direct}. For concreteness, we restrict attention in the main text to the randomized interventional analog to the natural direct effect ($rNDE$), which is identified under weaker conditions than the $NDE$; estimation algorithms for the $CDE$, $NDE$, $NIE$, and $rNIE$ are available in the supplemental materials. The $rNDE$ can be conceived as a contrast in dynamic regimes where part of the regime recreates the naturally occurring assignment mechanism \citep{didelez2012direct}. Because the naturally occurring assignment mechanism for $M$ is unknown, it must be estimated. We add a parametric model for $M$ reflecting its parents in the natural stochastic regime, and let $\theta = (\theta_Z, \theta_W, \theta_M, \theta_Y)$. The likelihood conditional on observed data is
\begin{equation*}
\mathcal{L}(\theta|o) = \mathcal{L}(\theta_Y | y, m, w, a, z)
\times 
\mathcal{L}(\theta_M|m, w, a, z)
\times
\mathcal{L}(\theta_W | w, a, z)
\times 
\mathcal{L}(\theta_Z | z). \label{eqn:medlike}
\end{equation*}
We now outline a strategy for estimating the $rNDE$ which compares $a=1$ to $a=0$ when the mediator is stochastically assigned as it would be under $a=0$. The $rNDE$ contrasts regime $g=(A=1, M=H_z(a=0))$ with $g'=(A=0, M=H_z(a=0))$, where $H_z(a=0)$ is a draw from the distribution of $M$ under $A=0$, conditional on $Z$. That is, $p(h_z(a=0)|\tilde{z}) = \int p(\tilde{m}|a=0, \tilde{z}, \tilde{w})p(\tilde{w}|a=0,\tilde{z})d\tilde{w}$.

The equation for $p(\tilde{y}^{g'} | o)$ for $g'=(A=0, M=H_z(a=0))$ is analogous to Equation~\ref{eqn:bgform} with an added model for $M$. For the regime $g=(A=1, M=H_z(a=0))$, the $W$ value used to assign $M$ is different from the value for $Y$ (i.e., a recanting witness) \citep{avin2005identifiability}. The simulation-based Bayesian g-formula resolves this issue with an independence assumption resulting in separate posterior predictive draws of $\tilde{w}^a$ for both $a=0$ and $a=1$. The posterior mean of the $rNDE$ is thus given by
\begin{align}
rNDE = \idotsint \bigg[ &
\tilde{y} 
\bigg( 
p(\tilde{y} | a=1, m, \tilde{w}^1, \tilde{z}, \theta_Y)
p(\tilde{w}^1 | a=1, \tilde{z}, \theta_W) - 
\nonumber \\
&
p(\tilde{y} | a=0, m, \tilde{w}^0, \tilde{z}, \theta_Y)
p(\tilde{w}^0 | a=0, \tilde{z}, \theta_W)\bigg) \times
\nonumber \\
&
p(\tilde{m} | a=0, \tilde{z}, \tilde{w}^0, \theta_M) 
p(\tilde{z} | \theta_Z)
\pi(\theta|o)
\bigg]
d\theta
d\tilde{w}^0
d\tilde{w}^1
d\tilde{m}
d\tilde{z}
d\tilde{y}. \label{eqn:bgformrnde}
\end{align}

\subsection{The Bayesian g-formula with the Bayesian bootstrap: an alternative marginalization strategy for closed-form estimands}
Previous sections assumed that marginalization over the baseline confounder distribution occurred through posterior predictive sampling from $p(\tilde{z}|\theta_Z)$. In practice, $Z$ can be high-dimensional, and parametrically modeling $p(z|\theta_Z)$ introduces additional opportunities for model misspecification. Because $Z$ is by definition the same for all regimes, \citet{keil2015bayesian} suggest sampling $\tilde{z}$ nonparametrically from the observed empirical distribution of $Z$, $p_N(z)$. For settings where $Z$, $A$, and $M$ are all discrete, we introduce an alternative approach. Because the g-formula estimators have closed forms in the discrete case, marginalization can occur through the Bayesian bootstrap. The Bayesian bootstrap assigns observation weights $(d_1, \dots, d_n)$ sampled from a $\mathrm{Dirichlet}(1,\dots,1)$, with weights changing every MCMC iteration. The closed-form causal contrast $\tau$ is calculated for every observed $Z$, yielding $\tau(Z_i)$ for $i=1,\dots,n$. Then the weighted average $\sum_{i=1}^n d_i \tau(Z_i)$ gives the posterior draw of the population average causal effect for that MCMC iteration. We refer to this procedure as the closed-form Bayesian g-formula (BGF-CF).

For relatively simple causal graphs with discrete data, BGF-CF avoids the computationally intensive posterior predictive simulation of BGF-SIM. Many popular causal estimands have tractable closed form solutions for discrete data. As an example, the $rNDE$ for baseline confounder value $z$, $rNDE(z)$, can be written as 
\begin{align}
\sum_{w, m} 
\big\{
&
\mathbb{E}\left[Y|a=1,w,m,z\right] p(w|a=1,z) 
- 
\mathbb{E}\left[Y|a=0,w,m,z\right] p(w|a=0,z)
\big\}
p(m|a=0,z). \label{eqn:rndegform}
\end{align}
Model-based estimates of every term in Equation~\ref{eqn:rndegform} can obtained with each posterior parameter sample $\theta^{(b)}$, with $p(m|a=0,z) = \sum_{w'} p(m|a=0,w',z,\theta_M^{(b)})p(w'|a=0,z,\theta_W^{(b)})$. If there are $K$ distinct baseline confounder patterns and $B$ MCMC iterations, the Bayesian g-formula only requires calculating $rNDE(z)$ a total of $B \times K$ times. Letting $\xi_k$ denote the sample frequency of covariate pattern $k$, the Bayesian bootstrap weights $(d_1,\dots,d_K)$ are repeatedly sampled from a $\mathrm{Dirichlet}(\xi_1,\dots,\xi_K)$. At each iteration, the posterior draw of the population $rNDE$ is $\sum_{k=1}^K d_k \times rNDE(z_k)^{(b)}$. For $K \ll n$, this method can be much more computationally efficient than the simulation-based g-formula. 

\section{Bayesian data fusion for unmeasured confounding}
\begin{figure}
		\centering 
		\begin{tikzpicture}[ ->,shorten >=2pt,>=stealth,node distance=1cm,pil/.style={->,thick,shorten =2pt}]
			\node (a) {$A$};
            \node[above left=of a] (z) {$Z$};
			\node[right=of a] (m) {$M$};
			\node[right=of m] (y) {$Y$};
			\node[below=of m] (c) {$U$};
            \draw[->] (z) to (a);
            \draw[->] (z) to (m);
            \draw[->] (z) to [out=10, in=120] (y);
            \draw[->] (z) to [out=270, in=180] (c);
			\draw[->] (a) to [out=30, in=150] (y);
			\draw[->] (a) to (m);
			\draw[->] (a) to (c);
			\draw[->] (m) to (y);
			\draw[->] (c) to (y);
			\draw[->] (c) to (m);
		\end{tikzpicture}
		\caption{Mediation causal structure with outcome $Y$, exposure $A$, mediator $M$, baseline confounders $Z$, and exposure-induced mediator-outcome confounder $U$ that is unmeasured in the main data}
		\label{fig:bdfmed}
\end{figure}
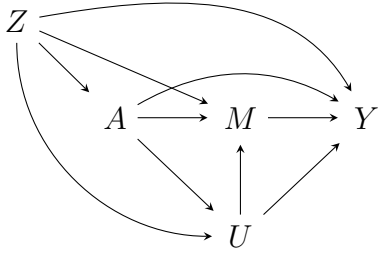

We now consider the problem of making causal inferences when an important confounder is unmeasured in the primary (``main'') data set but is available in a secondary (``external'') source. Although the data fusion algorithm we outline accommodates arbitrary confounding structures and many different estimands, we restrict attention to estimating a randomized natural direct effect with an exposure-induced unmeasured mediator-outcome confounder $U$ as in Figure~\ref{fig:bdfmed}. This setting is interesting for two reasons: (1) the additional complexity involved with estimating the natural, stochastic assignment mechanism and (2) the need to accommodate exposure-induced confounding. Complete estimation algorithms for the $rNIE$, $CDE$, and average treatment effects for longitudinal exposures are all available in the supplemental materials.

\subsection{Notation and assumptions}
Suppose that an investigator is interested in learning about an effect in some target population for which a large data source ($N=n_1$) exists. The desired causal quantity is the $rNDE$ in the population from which the $n_1$ observations were randomly sampled. Figure~\ref{fig:bdfmed} shows the causal structure, with outcome $Y$, exposure $A$, mediator $M$, and baseline confounders $Z$. There is also an exposure-induced mediator-outcome confounder $U$, which is unmeasured in the data set of size $n_1$. Information on $\{A,M,Z,Y,U\}$ exists in a smaller secondary data source ($N = n_2$, with $n_2 < n_1$), which may or may not be from the same population. For g-computation, models for $\{A, M, Y\}$ can be any univariate or multivariate generalized linear model, and there are no distributional restrictions on $Z$. However, imposing additional restrictions on $U$ can dramatically improve MCMC performance. Marginalization over the distribution of $U$ is only guaranteed for $U$ with finite support, although certain continuous distributions can also be integrated out of the likelihood.

In addition to the assumptions of the Bayesian g-formula \cite{keil2015bayesian} -- which are formally stated as Assumptions 1-4 in the Web Appendix -- Bayesian data fusion requires causal transportability to hold. That is, although the smaller data set may not be representative of the target population with respect to the distribution of baseline covariates, the underlying causal processes operate in the same way.
\setcounter{assumption}{4}
\begin{assumption}{(Parametric causal transportability)}
Let $\mathcal{P}_1$ and $\mathcal{P}_2$ denote the superpopulations for the main and external data sources. If $\mathcal{P}_1 \ne \mathcal{P}_2$, the causal graph structures of $\mathcal{P}_1$ and $\mathcal{P}_2$ must agree such that all $X \in \{U, M, Y\}$ have the same parent nodes $\mathrm{pa}(X)$. Furthermore, the true underlying data-generative parameters $\theta_{X, \mathcal{P}_1}$ and $\theta_{X, \mathcal{P}_2}$ must be the same such that
\begin{equation}
p(x | \mathrm{pa}(X), \theta_{X, \mathcal{P}_1}) = 
p(x | \mathrm{pa}(X), \theta_{X, \mathcal{P}_2}).
\end{equation}
Note that causal transportability holds by design if the external data are a random validation sample from $\mathcal{P}_1$.
\end{assumption}

The conditional exchangeability of Assumption 3 is still required for $U$ (i.e., $U^a \cindep A | Z$, a conditional independence which is encoded in Figure~\ref{fig:bdfmed}), but the requirements for $M$ and $Y$ are relaxed to allow the confounder $U$ to be unmeasured. 
\begin{assumption}{(Conditional exchangeability)}
Briefly, $(Z,U)$ must be sufficient to control confounding. For the randomized $rNDE$ in Figure~\ref{fig:bdfmed}, this implies:
\begin{align}
M^a \cindep A & | Z, U \\
Y^a \cindep A & | Z, U, M
\end{align}
\end{assumption}

\subsection{Specification of parametric models}\label{sec:bdfmodels}

To illustrate the closed-form estimator and facilitate contrasts with existing methods, we assume $A$, $U$, $M$, and $Y$ are all binary with logistic link functions. The baseline confounders $Z$ are also assumed to be discrete. 

Letting $\pi_X = P(X = 1| \mathrm{pa}(X))$, we adopt the following models: 
\begin{gather}
\logit{\piui} = 
\gamma_0 + \gamma_A A_i + z_i'\gamma_{Z} \label{eqn:bdfmodelu} \\
\logit{\pimi} = 
\beta_0 + \beta_A A_i + z_i'\beta_{Z} + \beta_U U_i \label{eqn:bdfmodelm} \\
\logit{\piyi} =  
\alpha_0 + \alpha_A A_i + z_i'\alpha_{Z} + \alpha_M M_i + \alpha_{AM}A_i M_i + \alpha_U U_i \label{eqn:bdfmodely}
\end{gather}
Then $\theta_U = (\gamma_0, \gamma_A, \gamma_Z)$, $\theta_M = (\beta_0, \dots, \beta_{U})$, and $\theta_Y = (\alpha_0, \dots, \alpha_U)$. For $X \in \{U, M, Y\}$, 
$\calL(\theta_X|x, \mathrm{pa}(X)) = f(x|\mathrm{pa}(X), \theta_X) = \prod_{i=1}^N (\pi_{X,i})^{x_i} (1-\pi_{X,i})^{1-x_i}$. Equation~\ref{eqn:bdfmarg} shows the observed data likelihood for the full parameter vector $\theta = (\theta_U, \theta_M, \theta_Y)$ in the main data set, marginalizing over the missing $U$.
\begin{align}
\mathcal{L}_m = \prod_{i=1}^{n_1} \left[ \sum_{u}
\mathcal{L}(\theta_Y | y_i, m_i, u_i=u, a_i, z_i)
\mathcal{L}(\theta_M|m_i, u_i=u, a_i, z_i)
\mathcal{L}(\theta_U|u_i=u, a_i, z_i)
\right] \label{eqn:bdfmarg}
\end{align}

For a generic prior $\pi(\theta)$, the posterior for $\theta$ marginalizing over the missing $U$ is proportional to $\mathcal{L}_m \times \pi(\theta)$. How to set an informative prior $\pi(\theta)$ using the secondary data set is the focus of the next section.

\subsection{Specification of prior information with external data} \label{sec:bdfprior}
Given that $U$ is unmeasured in the main data source, any parameters involving $U$ (i.e., $\theta_U$, $\beta_U$, and $\alpha_U$) cannot be identified from that data. Because the main data set is presumably more representative of the target population of interest, the sole reason for integrating the external data set is for providing information about the confounder unmeasured in the main data set. That information can be summarized through the use of informative priors.

Priors for $\{ \theta_U, \theta_M, \theta_Y \}$ are derived by fitting in the external data frequentist maximum likelihood models corresponding to Equations \ref{eqn:bdfmodelu} through \ref{eqn:bdfmodely}. Under causal transportability, maximum likelihood estimators fit in the external data will be consistent and asymptotically normal about $\theta_{\mathcal{P}_1}$. For $X\in\{U,M,Y\}$, let $\hat{\theta}_{X,MLE}$ denote the maximum likelihood estimate (MLE) of $\theta_X$ in the external data, and let $\widehat{\Sigma}_{X,MLE}$ be the estimated variance-covariance matrix. Then $\mathcal{N}(\hat{\theta}_{X,MLE}, \widehat{\Sigma}_{X,MLE})$ is a sensible choice for $\pi(\theta_X)$. With moderately large $n_2$, this prior approximates the posterior distribution for $\theta$ in a Bayesian analysis conducted using the $n_2$ observations, assuming a non-informative prior. With a priori independence, the complete prior for $\theta$ is  $\pi(\theta)=\pi(\theta_U)\times\pi(\theta_M)\times\pi(\theta_Y)$.

If $\mathcal{P}_1 \ne \mathcal{P}_2$, then less informative priors may be preferable for the identifiable parameters. Consider the parameter $\alpha_A$, about which the main data source contains substantial information. In the model formulation given by Equation~\ref{eqn:bdfmodely}, variance and covariance hyperparameters for $\alpha_A$ would be found along the second row and column of $\hat{\Sigma}_{Y, MLE}$. If we multiply the off-diagonal elements in that row and column by a large inflation factor (e.g., $\sigma=1000$) and the diagonal element by $\sigma^2$, we assert a marginal prior distribution on $\alpha_A$ that is virtually non-informative. However -- critically --  the prior correlation between $\alpha_A$ and the unidentifiable parameter $\alpha_U$ is preserved.

\subsection{A simulation-based Bayesian data fusion algorithm (BDF-SIM)}
We now outline a simulation-based Bayesian data fusion approach for g-formula causal contrasts in the context of $rNDE$ estimation.
\begin{enumerate}\label{alg:bdfsim}
\item Fit maximum likelihood models in the external data to obtain the prior $\pi(\theta)$ as detailed in Section~\ref{sec:bdfprior}.
\item Use a No-U-Turn sampler (NUTS) with target probability distribution proportional to $\mathcal{L}_m \times \pi(\theta)$ in order to obtain posterior samples of the regression parameter vector $\theta$. The probabilistic programming language Stan has a NUTS implementation \citep{carpenter2016stan}, and it is available to R users through the \texttt{rstan} R package \citep{rstan}. For some large $B$ (e.g., 4,000), let  $\theta^{(1)}, \dots, \theta^{(B)}$ denote the $B$ posterior samples remaining after discarding warmup iterations.
\item For MCMC iteration $b=1,\dots,B$ and $i = 1,\dots, n_1$:
\begin{enumerate}[a)]
\item Sample baseline covariate vector $\tilde{z}_i$ from the empirical distribution.
\item For $g_0 \in \{ g, g' \}$ and $a_0 \in \{ 0 , 1 \}$, sample $\tilde{u}^{a_0, g_0 (b)}_i$ as Bernoulli with success probability 
\[ \logitinv{ \gamma_0^{(b)} + \tilde{z}_i' \gamma_Z^{(b)} + a_0 \gamma_A^{(b)} } \]
\item For $g_0 \in \{ g, g' \}$, sample randomized mediator $\tilde{m}^{0, g_0 (b)}_{i}$ as Bernoulli with success probability
\[ \logitinv{ \beta_0^{(b)} + \tilde{z}_i' \beta_Z^{(b)} + \beta_U^{(b)} \tilde{u}^{0, g_0 (b)}_{i} } \]
\item Define individual-level causal contrast $\tilde{\phi}_i^{(b)}$ as 
\begin{align*}
\tilde{\phi}_i^{(b)} = & 
\logitinv{\alpha_0^{(b)} + \tilde{z}_i' \alpha_Z^{(b)} + \alpha_M^{(b)} \tilde{m}^{0, g (b)}_{i} + \alpha_A^{(b)} + \alpha_{AM}^{(b)} \tilde{m}^{0, g (b)}_{i} +
\alpha_U^{(b)} \tilde{u}^{1, g (b)}_{i} } \\
& -  
\logitinv{\alpha_0^{(b)} + \tilde{z}_i' \alpha_Z^{(b)} + \alpha_M^{(b)} \tilde{m}^{0, g' (b)}_{i} +
\alpha_U^{(b)} \tilde{u}^{0, g' (b)}_{i} }
%
\end{align*}
\end{enumerate}
\item Calculate population estimate $rNDE^{(b)} = \sum_{i=1}^{n_1} \tilde{\phi}_i^{(b)} / n_1$.
\item Construct a point estimate for $rNDE$ as the posterior mean $\widehat{rNDE}=\sum_{b=1}^{B} rNDE^{(b)}/B$, and create quantile-based 95\% credible intervals as the $2.5^{th}$ and $97.5^{th}$ quantiles of $(rNDE^{(1)}, \dots, rNDE^{(B)})$.
\end{enumerate}

\subsection{A Bayesian data fusion algorithm for closed-form estimands using the Bayesian bootstrap (BDF-CF)}\label{sec:alg:gform}
We now outline a data fusion procedure for the $rNDE$ using the Bayesian g-formula with the Bayesian bootstrap for confounder marginalization. Steps 1, 2, and 5 of the closed-form version with the Bayesian bootstrap are identical to the simulation-based approach, so we show only steps 3 and 4.

\begin{enumerate}[1.]\setcounter{enumi}{2}
\item For $b=1,\dots,B$ and $k=1,\dots,K$ for the $K$ unique covariate patterns
\begin{enumerate}[a)]
\item Sample covariate patterns weights $(d_1^{(b)}, \dots, d_K^{(b)})$ from a $\mathrm{Dirichlet}(\xi_1,\dots,\xi_K)$, where $\xi_k$ is the count of observations with the unique covariate pattern $Z_k$.
\item Calculate the individual-level contrast $\phi(z_k)$ for pattern $z_k$ according to Equation~\ref{eqn:bgformrnde}, replacing $W$ with $U$ and plugging in the appropriate model-based estimates from $\theta^{(b)}$. Concretely,
\begin{align*}
\mathbb{E}\left[Y|a,u,m,z\right] 
= & 
\logitinv{\alpha_0^{(b)} + z' \alpha_Z^{(b)} + \alpha_M^{(b)} m + \alpha_A^{(b)} a +
\alpha_{AM}^{(b)} am +
\alpha_U^{(b)} u } \\
p(u|a,z) 
= & 
\logitinv{ \gamma_0^{(b)} + \gamma_A^{(b)} a + z' \gamma_Z^{(b)} } \\
p(m|a=0,z)
= &
\sum_{u} \left(
\logitinv{ \beta_0^{(b)} + z' \beta_Z^{(b)} + \beta_U^{(b)} u} p(u|a=0,z) \right)
\end{align*}
\end{enumerate}
\item Calculate population estimate $rNDE^{(b)} = \sum_{k=1}^K (d_k^{(b)} \times \phi_k^{(b)})/n_1$.
\end{enumerate}

Although the simulation-based and closed-form Bayesian g-formula approaches are identical with respect to regression parameter estimation, their differences have implications for extensibility to other causal estimands and scalability to large data sets. The performance of these two estimators under various conditions is the focus of the next section.

\section{Simulation study}\label{sec:sims}

We designed a simulation study to evaluate the performance of the simulation-based and closed-form BDF estimators relative to existing bias corrections we briefly describe in this section. The estimand of interest was the $rNDE$ in the main study superpopulation. Due to the fact that sensitivity analyses based on sharp nonparametric bounding and those based on externally derived bias parameters are not directly comparable, we focus our comparison between BDF and other bias correction techniques.
 
\subsection{Data generation procedure}
We considered a number of scenarios with varying data generation schemes. We varied: sample sizes ($n_1$ and $n_2 =n_1/10$), causal structure ($\Delta_{U,A}$ = 1 if the mediator-outcome confounder $U$ is caused by $A$ and 0 otherwise), and presence of an interaction ($\Delta_{Y,AM}$ = 1 if there is an $A\text{-}M$ interaction in the $Y$ model and 0 otherwise). The strength of mediator-outcome confounding by $U$ was governed by two quantities, $\beta_U$ and $\alpha_U$, the log-odds ratios of $U$ in the $M$ and $Y$ models, respectively. When the same $\beta_U$ and $\alpha_U$ were used to generate the main and external data sets, we have causal transportability; this was done for strong confounding by $U$ ($\beta_U = \alpha_U = 1.5$, corresponding to odds ratios of $\approx 4.5$). To investigate the performances of the various approaches under violations of the transportability assumption, $\beta_U = \alpha_U = 0$ was used to generate the external data, while $\beta_U = \alpha_U = 1.5$ in the main data. Complete details of the data generation process are available in the Web Appendix. For each simulation condition, estimator bias and interval coverage were assessed using 200 replicates.

\subsection{Implementation of the Bayesian data fusion estimators}
Closed-form and simulation-based variants of Bayesian data fusion were implemented for each pair of simulated main and external data sets. Priors were constructed using the external data as described in Section~\ref{sec:bdfprior} without variance inflation for the covariance matrix (i.e., assuming transportability). The models in the external data were correctly specified, with $\Delta_{Y,AM}$ and $\Delta_{U,A}$ matching the underlying generation process for the main data. Posterior samples of the bias-corrected $rNDE$ were obtained from 3 MCMC chains of 2,000 iterations each, with the first 1,000 samples discarded as warmup. The posterior mean was taken as a point estimate, with uncertainty captured using 95\% quantile-based credible intervals.

\subsection{Alternative bias correction methods}
The first comparator method, referred to as the delta-gamma (DG) correction, is a classical bias correction method \citep{vanderweele2015explanation}. A version for controlled direct effects can be used for the $rNDE$ when the two coincide, i.e., if (1) $U$ is not exposure-induced, and (2) there is no exposure-mediator interaction in the outcome model. Note that for estimands on the risk difference scale, (2) does not hold for logistic models of $Y$ even if the $A\text{-}M$ interaction coefficient is zero. This approach also requires that the effect of $U$ should be the same across all levels of $A$ (i.e., $\E{Y|a,z,m,U=1} - \E{Y|a,z,m,U=0}$ does not depend on $a$), which cannot hold in a logistic model unless the $A$ coefficient is zero. For comparability with BDF we elected to use the secondary data source, replacing component quantities in the DG bias formula with maximum likelihood estimates derived from logistic regression models in the external data. Confidence intervals were obtained by bootstrapping the main data 200 times. 

A second frequentist correction, referred to as the interaction correction (IX), can accommodate exposure-mediator interaction in the outcome model \citep{vanderweele2010bias}. Originally derived as a bias correction for the $NDE$, it is more generally applicable than the DG correction, but it similarly requires that $U$ not be exposure-induced. Again, we fit maximum likelihood models to the external data source to derive bias-corrected estimates within each covariate pattern, and confidence intervals were obtained via the bootstrap. Additional details on the DG and IX  implementation are available in the supplemental materials.

\subsection{Simulation results}
\begin{figure}[ht!]
    \centering
    \includegraphics[width = 0.95\textwidth]{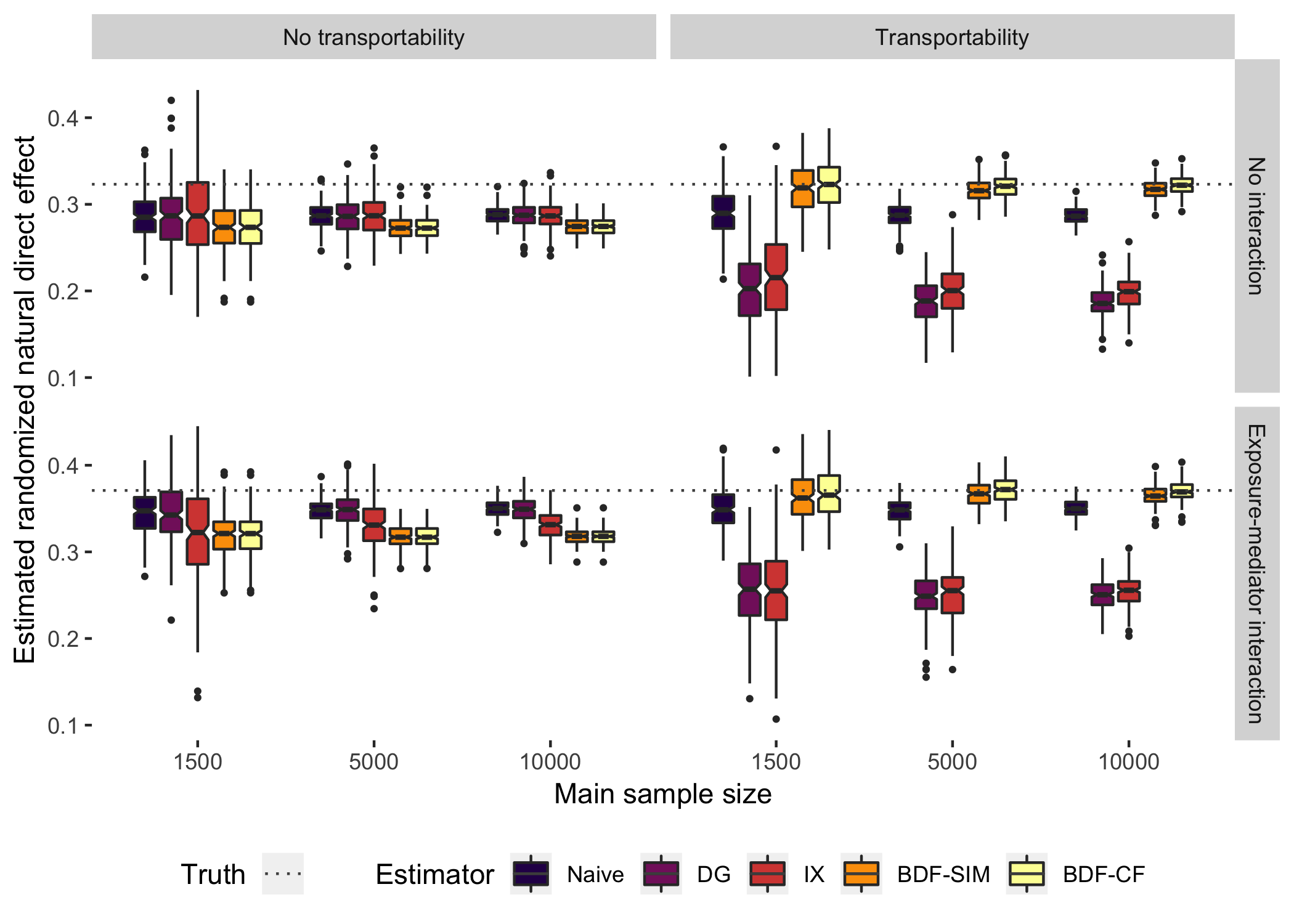}
    \caption{Randomized natural direct effects estimated with naive, delta-gamma (DG) correction, interaction (IX) correction, simulation-based Bayesian data fusion (BDF-SIM), and closed-form Bayesian data fusion (BDF-CF) estimators, with and without exposure-mediator interaction and causal transportability between main and external data sets.}
    \label{fig:simbox}
\end{figure}

Figure~\ref{fig:simbox} shows estimates from the case where $U$ is exposure-induced. When the transportability assumption holds, both BDF estimators eliminate the confounding bias at all sample sizes. In contrast, the frequentist correction methods do worse than no correction at all. Although these corrections do not purport to address exposure-induced mediator-outcome confounding, this finding underscores the danger of using these corrections when $U$ may be caused by $A$.

In the absence of transportability, the information extracted from the external data set by the BDF procedure is misleading, and the estimators perform poorly. Confounding bias is not eliminated, and the prior information leads to less posterior uncertainty. The frequentist estimators also do not correct the bias, but the uncertainty is the same as the uncorrected naive intervals.

Table~\ref{tab:cov} reports coverage for the 95\% confidence and credible intervals; the Web Appendix contains information regarding interval widths. In general, credible intervals from BDF approaches had widths comparable to the naive and frequentist corrected confidence intervals. However, the classical correction methods are not unbiased, and the interval coverage is low ($< 10\%$). Conversely, as noted previously, BDF methods did not perform well in the case of transportability violations, achieving less than 1\% coverage.

\begin{table}
\caption{\label{tab:cov}Coverage percentages for 95\% confidence and credible intervals for
             naive, delta-gamma (DG) and interaction (IX) frequentist corrections, 
             simulation-based (BDF-SIM) and closed-form (BDF-CF) Bayesian data fusion estimators, 
             calculated in 200 replicates with exposure-induced mediator-outcome confounding}
\centering
\begin{tabular}[t]{lllrrrrr}
\toprule
Transportability & Interaction & Sample sizes & Naive & DG & IX & BDF-SIM & BDF-CF\\
\midrule
Yes & No & (150, 1500) & 73.5 & 5.0 & 15.5 & 93.5 & 92.5\\
Yes & No & (500, 5000) & 25.0 & 0.0 & 0.0 & 94.0 & 95.0\\
Yes & No & (1000, 10000) & 2.5 & 0.0 & 0.0 & 91.5 & 94.5\\
Yes & Yes & (150, 1500) & 69.0 & 7.0 & 12.5 & 92.5 & 91.5\\
Yes & Yes & (500, 5000) & 19.5 & 0.0 & 0.0 & 95.5 & 96.0\\
Yes & Yes & (1000, 10000) & 2.5 & 0.0 & 0.0 & 93.5 & 94.5\\
\addlinespace
No & No & (150, 1500) & 66.5 & 64.0 & 54.0 & 50.0 & 50.0\\
No & No & (500, 5000) & 26.0 & 31.5 & 34.0 & 3.0 & 2.5\\
No & No & (1000, 10000) & 2.5 & 10.5 & 13.0 & 0.0 & 0.0\\
No & Yes & (150, 1500) & 59.5 & 72.5 & 45.0 & 53.0 & 49.5\\
No & Yes & (500, 5000) & 18.0 & 55.0 & 33.0 & 3.5 & 3.5\\
No & Yes & (1000, 10000) & 3.5 & 44.0 & 13.0 & 0.0 & 0.0\\
\bottomrule
\end{tabular}
\end{table}

\section{Examining the role of stage at diagnosis in Black-White survival disparities in colorectal cancer}\label{sec:dataapp}
\subsection{Overview}
We now use BDF to explore the extent to which differentials in stage at diagnosis contribute to apparent racial disparities in colorectal cancer survival. Our analysis provides an estimate of how much we could reasonable expect to reduce the observed survival disparity if we could break the  between race and cancer stage at the time of diagnosis, e.g., by implementing targeted screening programs that lead to earlier colorectal cancer detection among Blacks.

\citet{valeri2016role} sought to address this question in a recent article with data from a registry of US cancer patients from 1992-2005. The National Cancer Institute's Surveillance, Epidemiology, and End Results (SEER) registry collects information on tumor site and stage for a sizable proportion of cancer patients from diverse geographic regions within the US. They concluded that eliminating Black-White disparities in colorectal cancer stage at diagnosis would lead to a 35\% reduction in survival disparities as measured by 5-year restricted mean survival time. Their analysis controlled for a number of covariates, including age at diagnosis, gender, time period of cancer diagnosis, geographic locale, and median county income as derived from the American Community Survey; however, it did not control for household-level poverty status, as that information was not available.

\subsection{Analysis description}
We extend the analysis of Valeri and colleagues by incorporating information about confounding of the stage-survival relationship by individual-level income using data from the Cancer Care Outcomes Research and Surveillance (CanCORS) Consortium data. This observational study followed patients shortly after cancer diagnosis and aimed ``to determine how the characteristics and beliefs of lung and colorectal cancer patients, physicians and health-care organizations influence treatments and outcomes spanning the continuum of cancer care from diagnosis to recovery or death, and to evaluate the effects of specific therapies on patients’ survival, quality of life, and satisfaction with care'' \citep{catalano2013representativeness}. As a result of these ambitious aims, the CanCORS database contains detailed socioeconomic information, including household income for the year preceding cancer diagnosis. We chose $U=1$ to correspond the lowest income group of $<$\$40,000 per year. The goal was to assess the bias of the residual disparity measure as calculated in SEER, assuming true underlying race-poverty and poverty-survival relationships in SEER matched those estimated in CanCORS. The survival outcome was a binary indicator $Y$ for whether the patient was alive 5 years post-cancer diagnosis. Self-reported race was coded such that $A=1$ for non-Hispanic blacks and $A=0$ for non-Hispanic whites; individuals reporting Hispanic origin were excluded. The intervening variable of interest, stage at cancer diagnosis $M$, took on values 1-4 corresponding to cancer stages I-IV. Adjustment covariates included in all models were: gender, age at cancer diagnosis ($<$60, 60-65, or $>$65), and geographic region (West, South, or other). Patients whose cancer was unstaged were excluded, leaving a total of 146,031 colorectal cancer cases in the SEER analysis data set.

First, we fit two naive models using maximum likelihood in the SEER data: (1) stage at cancer diagnosis as a function of race and adjustment covariates, using a baseline category logit model; and (2) 5-year survival as a function of race, stage at diagnosis, and the adjustment covariates, using a logistic link. Coefficients from these models were used to calculate a naive estimated residual disparity measure $RD_{naive}$ and bootstrapped 95\% confidence intervals.

Next, we implemented both BDF estimators to obtain poverty-adjusted estimates of the black-white survival disparity. To construct priors, we fit three frequentist models using the 1,613 CanCORS colorectal cancer patients for whom complete stage and covariate data were available. The two regression models described above were modified by adding poverty as a covariate. Since SEER is more representative of the target population for intervention, all parameters except the bias coefficients were given marginally noninformative prior distributions with the variance inflation strategy outlined in Section~\ref{sec:bdfprior}. A third and final frequentist model was a logistic regression for poverty as a function of race, gender, region, and age category. Because none of these $U$-related parameters are identifiable in SEER, no variance inflation procedure was performed.

Using simulation and closed-form BDF, we estimated the poverty-adjusted residual disparity in the SEER data. Four chains of 2,000 MCMC iterations each were run in Stan \citep{rstan}, with the first 1,000 iterations discarded as warmup. The Gelman-Rubin convergence diagnostic $\hat{R}$ was calculated for all parameters \citep{gelman1992inference}.

\subsection{Residual disparity results}
\begin{figure}
\centering
\includegraphics[width = 0.95\textwidth]{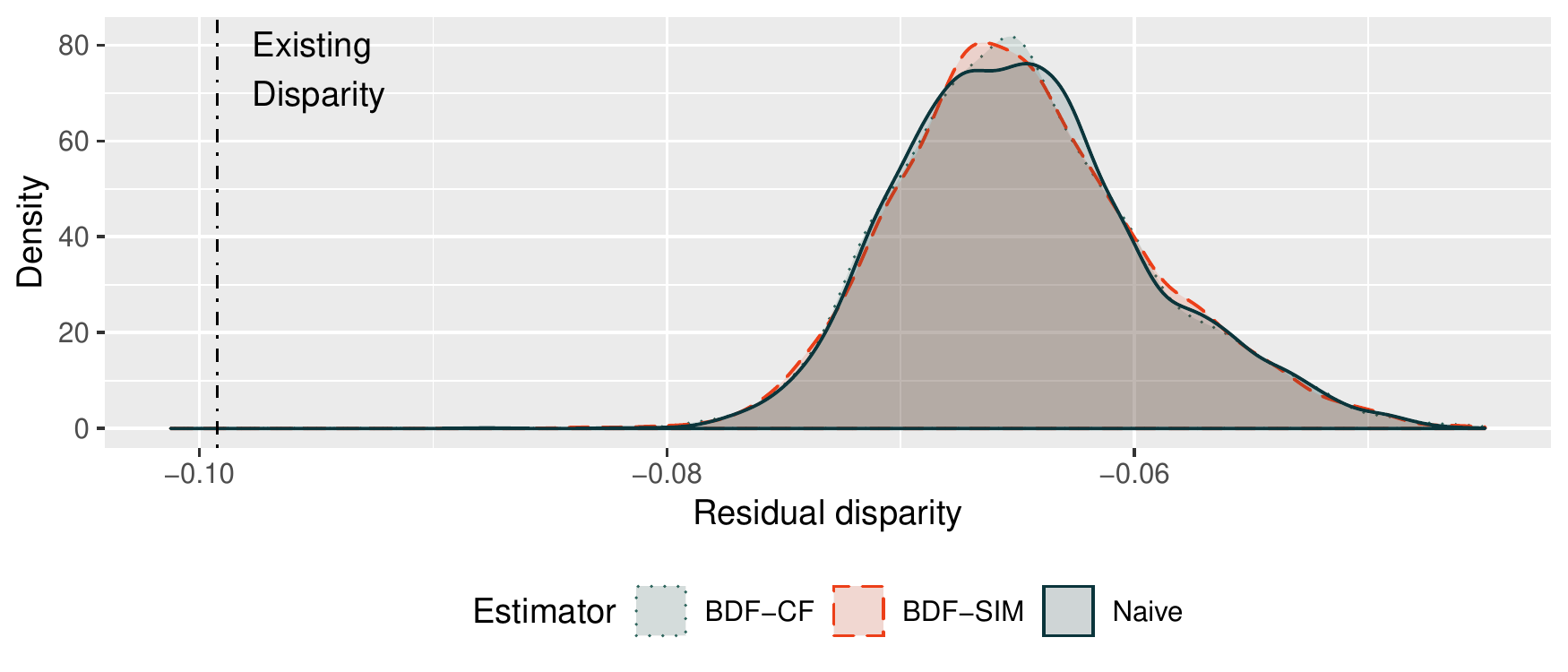}
\caption{Posterior samples of average residual disparity (ARD) estimates of differences in Black-White 5-year colorectal cancer survival probabilities in the Surveillance Epidemiology and End Results (SEER) population, accounting for unmeasured poverty using closed-form (BDF-CF) and simulated (BDF-SIM) Bayesian data fusion from the CanCORS cohort study}
\label{fig:dataappdensities}
\end{figure}

Posterior samples of the poverty-adjusted population residual disparity measure calculated using simulation-based and closed-form BDF are shown in Figure~\ref{fig:dataappdensities}. The null value of zero, which represents Black-White equality with respect to baseline covariate-adjusted survival, lies beyond the far right of the graph. Visible as a dotted line on the left is the disparity we currently observe without an intervention on stage. With a value of 0.099 (95\% CI: 0.092, 0.107), we estimate that Black patients are 9.9 percentage points less likely to survive 5 years post-diagnosis than White patients of the same gender and geographic region. The naive estimate of the residual disparity after an intervention aligning Blacks' cancer stage distribution to the current stage suggests that the remaining disparity in 5-year survival would be 6.6\% (95\% CI: 5.8, 7.4). The BDF analyses suggest that unmeasured confounding by poverty does not substantially change the estimated residual disparity, with closed-form and simulation-based estimates of 6.5\% (95\% CI: 5.3, 7.4).

Given the abundance of literature documenting the role of socioeconomic status in cancer outcomes \citep{le2008effects}, it may be surprising to see adjustment for poverty having such a small impact on the estimated residual disparity. One possibility is that causal transportability may not hold between the SEER and CanCORS populations in ways related to poverty (i.e., which cannot be addressed with variance inflation). That is, CanCORS may appear to be representative of the larger US population from which SEER draws its cancer cases \citep{catalano2013representativeness}, but the causal relationships determining cancer outcomes in CanCORS are fundamentally different from the processes in SEER because many CanCORS study sites are academic medical centers in large cities \citep{ayanian2004understanding}. Thus, we may not see dramatic shifts in our conclusions for SEER because CanCORS does not contain evidence for substantial stage-survival confounding by poverty. Alternatively, there may residual confounding due to the coarsening of socioeconomic deprivation -- a complex, multifaceted problem -- into a single binary indicator. Nevertheless, this analysis integrating the two data sources gives policymakers two potentially valuable pieces of information: (1) a quantitative estimate of the poverty-adjusted residual disparity and (2) a better understanding of the true uncertainty surrounding that estimate. 

\section{Discussion}\label{sec:disc}
In this paper, we have proposed a general method for Bayesian data fusion that can be used to perform sensitivity analyses for unmeasured confounding in a variety of settings. The approach addresses forms of confounding in both static and dynamic treatment regimes as well as in mediation, including exposure-induced mediator-outcome confounding. While there can be no substitute for a well-designed study in the target population of interest, decision makers cannot wait for the ideal analysis in the ideal data set and must often work from incomplete or imperfect information. Bayesian data fusion communicates the sensitivity of a research conclusion while incorporating some of what is already known about the problem.

The general nature of the BDF-SIM and BDF-CF data fusion methods make them easily extendable to other settings. For example, any number of parametric models could be used for the unmeasured confounder. Although we demonstrated properties using the randomized natural direct effect in a mediation setting, these principles can be applied to any mediation estimand or to settings with time-varying confounding. With BDF-SIM, any generalized linear model can be adopted for the unmeasured confounder, allowing for both continuous and discrete distributions. Both BDF-SIM and BDF-CF can accommodate multiple unmeasured confounders. Depending on the types of confounding present, information on multiple confounders could be constructed from different external data sources, although doing so may require some assumptions about the joint distribution of their effects in the outcome model. 

With respect to the motivating question of Black-White racial disparities in U.S. colorectal cancer patients, we conclude that unmeasured confounding of the stage-survival relationship by poverty leads to residual disparity reduction estimates that are slightly too optimistic. Implementing an intervention -- for example, a targeted screening program -- to alleviate or eliminate delayed cancer diagnosis for Black colorectal cancer patients would substantially improve 5-year survival outcomes. However, without also intervening upon the complex societal factors that lead to greater poverty among black patients, we cannot realize the full benefit of such an intervention.

Both the closed-form and simulation-based variations of BDF are limited in part by their reliance on two major assumptions. First, they assume that the parametric models are otherwise correctly specified, which means they do not account for uncertainty in model misspecification, as some sensitivity analyses do \citep{tchetgen2012semiparametric}. A more comprehensive uncertainty quantification would incorporate additional uncertainty due to model selection. Second, as with most parametric causal inference, the models extrapolate causal effects \citep{Vansteelandt2012uncertainty}, and problems with overlap must be detected by the analyst. Recent advances in Bayesian nonparametrics \citep{roy2017bayesian} may be adapted to add flexibility to portions of the models. Third, like other data fusion methods, BDF assumes causal transportability between the external and main data sets \citep{pearl2014external}. We can perform exploratory analyses investigating representativeness, but we can never be certain that the underlying causal generative processes in two study populations are comparable. With BDF, the skeptical analyst can modify the variance-covariance matrices in the prior distribution to have greater variances, effectively reducing the ``prior sample size'' of the information on the unmeasured confounder. However, doing so places greater prior probability mass on confounding parameter values that may be implausible. Taken to the extreme, the external data source provides no information at all. When it comes to transportability and data fusion, one must balance caution about non-transportability and the desire to let the secondary data source inform the choice of sensitivity parameters. Thus, for data-driven sensitivity analysis methods to gain widespread acceptance, the statistical literature needs more formal methods for evaluating and characterizing causal transportability.

Nevertheless, these BDF methods represent significant steps forward in statistical analyses to inform decisionmaking. As a sensitivity analysis method, the flexible nature of BDF-SIM allows for a much greater variety of causal structures and regression model specifications, and it works for any causal estimand that can be represented as a counterfactual contrast. As a data fusion approach, the underlying Bayesian principles allow for extensibility to handle uncertainty quantification for multiple unmeasured confounders of various types. Information from the external data source enters exclusively through prior distributions, reducing computational burden and sidestepping data privacy concerns. BDF only needs  parameter estimates and variance-covariance matrices from the external data set; sharing this information does not compromise the privacy of individual study participants. This fact dramatically increases the number of data sources which may be used as external data. Researchers in the era of Big Data cannot guarantee that the right data are always available to them, but developing statistical methods to rigorously synthesize information from multiple sources can give decision-makers the tools to make more informed choices.


\section*{Acknowledgements}
LC acknowledges support from NIH grants T32CA009337 and T32ES007142 as well as a David H. Peipers Fellowship. CZ was supported by EPA grant RD-835872 and NIH grants R01ES026217 and NIH R01GM111339. LV received funding from NIH grant K01MH118477, and BC was funded by NIH grants P01CA134294 and P30ES000002.



\appendix
\section*{Supplemental materials}

\section{Assumptions for the Bayesian g-formula}
\label{sec:bgfassumptions}
\setcounter{assumption}{0}
\begin{assumption}[Positivity]
Let $X$ be any node in the causal graph $G$, and $x$ any value in the support of $X$. Then for any regime $g_0 \in \{ g, g'\}$, $p(x| \mathrm{pa}_{g_0}(X) = \tilde{pa}) > 0$, where $\tilde{pa}$ is any value taken by $\mathrm{pa}_{g_0}(X)$, the parent nodes of $X$ under $g_0$. Furthermore, it must hold for all $x$ and $\tilde{pa}$ that $p(x| \mathrm{pa}(X) = \tilde{pa}) > 0$, where $\mathrm{pa}(X)$ without a subscript indicates the parent nodes of $X$ in the naturally occurring treatment assignment mechanism.
\end{assumption}
\begin{assumption}[Consistency]
For any regime $g_0 \in \{g, g'\}$, $Y = Y^{g_0}$ whenever $V$ takes on the values prescribed by $g_0$. If $V$ is a single binary treatment, this statement simplifies to $Y = V Y^1 + (1 - V)Y^0$.
\end{assumption}
\begin{assumption}[Conditional exchangeability]
For any variable $V_0$ in the intervention set and every regime $g_0 \in \{g, g' \}$, there exists a set of measured variables $C \subset \{Z, W\}$ such that $Y^{g_0} \cindep V_0 | C$. 
\label{ass:strongce}
\end{assumption}
\begin{assumption}[Correct parametric model specification]
For every node $X \in \{V, W, Y\}$ modeled conditional on variables $C$ with parameters $\theta_X$, the parametric model $f(X | C,  \theta_X)$ is correctly specified.
\end{assumption}
 
\section{Bayesian g-formula algorithms for other causal estimands}
For simplicity of exposition, we assume a single unmeasured confounder $U$ throughout the following algorithms, but $U$ can be vector-valued, or there may be $L$ distinct unmeasured confounders $U_1, U_2, \dots, U_{L}$ occupying different positions in the causal graph. Although the algorithms outlined below are the simulation-based BDF-SIM, the more computationally efficient BDF-CF is available when all variables are discrete and in selected other instances (e.g., certain outcome models with identity link functions). Throughout, we continue using $n_1$ to denote the sample size of the main data sources.

\subsection{Time-varying confounding of a longitudinal exposure}
\begin{figure}
		\centering 
		\begin{tikzpicture}[ ->,shorten >=2pt,>=stealth,node distance=1cm,pil/.style={->,thick,shorten =2pt}]
			\node (a) {$A_1$};
            \node[above left=of a] (z) {$Z$};
			\node[right=of a] (m) {$A_2$};
			\node[right=of m] (y) {$Y$};
			\node[below=of m] (c) {$U$};
            \draw[->] (z) to (a);
            \draw[->] (z) to (m);
            \draw[->] (z) to [out=10, in=120] (y);
            \draw[->] (z) to [out=270, in=180] (c);
			\draw[->] (a) to [out=30, in=150] (y);
			\draw[->] (a) to (m);
			\draw[->] (a) to (c);
			\draw[->] (m) to (y);
			\draw[->] (c) to (y);
			\draw[->] (c) to (m);
		\end{tikzpicture}
		\caption{Time-varying causal structure with outcome $Y$, exposures $A_1$ and $A_2$, baseline confounder(s) $Z$, and time-varying confounder $U$}
		\label{fig:tvcdag}
\end{figure}
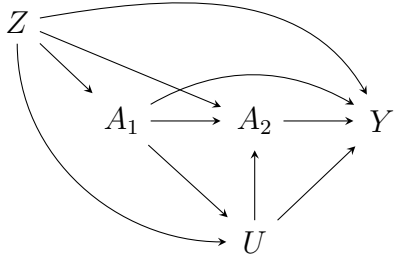

Suppose that the true causal DAG is as in Figure~\ref{fig:tvcdag}, with a discrete exposure-induced unmeasured variable $U$ acting as a confounder of the time-varying exposure $A$ measured at two time points to yield $A = (A_1, A_2)$. Denote the regimes of interest with $g = (A_1 = a_1, A_2 = a_2)$ and $g' = (A_1 = a_1', A_2 = a_2')$, with the causal estimand of interest being the superpopulation average causal effect $ACE = \E{Y^g - Y^{g'}}$.

Adopt parametric generalized linear models indexed by $\theta_X$ for $X \in \{U, A_2, Y\}$, with $h_X(\cdot)$ denoting the link function and $\eta_{X}$ the linear predictor term, which is a function only of the parent nodes $\pa{X}$ and $\theta_X$. Equation~\ref{eq:glmtvc} gives a general model form. 
\begin{align}
h_X(X_i | \mathrm{pa}(X_i), \theta_X) = & \eta_{Xi} \label{eq:glmtvc}
\end{align}
The intervention set is $V = (A_1, A_2)$. For $g_0 \in \{g, g'\}$ let $\mathrm{pa}^{g_0}(X)$ be the intervened parents of $X$ where any $V \in  \{ \mathrm{pa}(X) \}$ has been set (deterministically or stochastically) in accordance with $g_0$ and the relevant model.

To emphasize that $\eta_{Xi}$ depends on the parameters $\theta_X$ as well as the values of the parents of $X$ for observation $i$, we can also write it as $\eta_{Xi}(\mathrm{pa}(X), \theta_X)$. 

Let the  $\mathcal{L}(\theta_X | x , \mathrm{pa}(X))$.
Then $\theta = (\theta_U, \theta_{A_2}, \theta_Y)$ and
\begin{align}
\mathcal{L}_c = & 
\prod_{i=1}^{n_1} 
\mathcal{L}(\theta_Y | y_i, a_{2i}, u_i, a_{1i}, z_i)
\mathcal{L}(\theta_{A_2} | a_{2i}, u_i, a_{1i}, z_i)
\mathcal{L}(\theta_{U} | u_i, a_{1i}, z_i)
\end{align}
For discrete $U$, this yields the marginal likelihood of
\begin{align}
\mathcal{L}_m = & 
\prod_{i=1}^{n_1} \left[ \sum_u
\mathcal{L}(\theta_Y | y_i, a_{2i}, u_i = u, a_{1i}, z_i)
\mathcal{L}(\theta_{A_2} | a_{2i}, u_i = u, a_{1i}, z_i)
\mathcal{L}(\theta_{U} | u_i = u, a_{1i}, z_i) \right]
\end{align}

\begin{enumerate}\label{alg:bdftvc}
\item Fit maximum likelihood models in the external data to obtain the prior $\pi(\theta)$ as detailed in Section~\ref{sec:bdfprior}.
\item Use NUTS with target probability distribution proportional to $\mathcal{L}_m \times \pi(\theta)$ in order to obtain posterior samples of the regression parameter vector $\theta$. For some large $B$, let  $\theta^{(1)}, \dots, \theta^{(B)}$ denote the $B$ posterior samples remaining after discarding warmup iterations.
\item For MCMC iteration $b=1,\dots,B$ and $i = 1,\dots, n_1$:
\begin{enumerate}[a)]
\item Sample baseline covariate vector $\tilde{z}^{(b)}_i$ from the empirical distribution.
\item For $g_0 \in \{g, g' \}$, set $\tilde{a_1}^{g_0 (b)}_i$ deterministically or stochastically in accordance with $g_0$. For example, if $g$ is the static, deterministic regime setting $A_1$ to level $a_1$, $\tilde{a_1}^{g (b)}_i = a_1$ for all $i$ and $b$.
\item For $g_0 \in \{g, g' \}$, sample $\tilde{u}^{g_0 (b)}_i$ in accordance with
\[ h_X^{-1}\left( \eta_{X}\left( \mathrm{pa}^{g_0}(\tilde{x}^{g_0}_i), \theta_X^{(b)}\right) \right) \]
Concretely, for $g = (A_1 = a_1, A_2 = a_2)$ and a logistic model $\logit{P(U_i = 1 | Z_i, A_{1i})} = \gamma_0 + \gamma_{A_1} A_{1i} + \gamma_{Z}'Z_i$, sampling $\tilde{u}^{g (b)}_i$ requires drawing from a Bernoulli with success probability
\[ \logitinv{ \gamma_0^{(b)} + \gamma_{A_1} a_1 + \gamma_{Z}^{(b)\prime} \tilde{z}^{(b)}_i } \]
\item For $g_0 \in \{g, g' \}$, set $\tilde{a_2}^{g_0 (b)}_i$ deterministically or stochastically in accordance with $g_0$. Concretely, if $g$ is the static, deterministic regime setting $A_2$ to level $a_2$, $\tilde{a_2}^{g (b)}_i = a_2$ for all $i$ and $b$.
\item For $g_0 \in \{g, g' \}$, draw $\tilde{y}^{g_0 (b)}_i$ in accordance with $g_0$, $\theta_Y^{(b)}$, $\tilde{z}^{(b)}_i$, $\tilde{u}^{g_0 (b)}_i$, $\tilde{a_1}^{g_0 (b)}_i$, and $\tilde{a_2}^{g_0 (b)}_i$. Then calculate the individual-level causal effect
\begin{align*}
\tilde{\phi}_i^{(b)} = & \tilde{y}^{g (b)}_i - \tilde{y}^{g' (b)}_i
\end{align*}
Alternatively, if the conditional mean of $Y$ has a closed form $\mu(\theta_Y, z, u, a_1, a_2)$, define the individual-level causal effect as
\begin{align*}
\tilde{\phi}_i^{(b)} = & 
\mu\left(\theta_Y^{(b)}, \tilde{z}^{(b)}_i, \tilde{u}^{g (b)}_i, \tilde{a_1}^{g (b)}_i, \tilde{a_2}^{g (b)}_i \right) - 
\mu\left(\theta_Y^{(b)}, \tilde{z}^{(b)}_i, \tilde{u}^{g' (b)}_i, \tilde{a_1}^{g' (b)}_i, \tilde{a_2}^{g' (b)}_i \right).
\end{align*}
For example, if $Y_i$ is conditionally normal with mean $\alpha_0 + \alpha_{Z}' Z_{i} + \alpha_{A_1} A_{1i} + \alpha_{U} U_i + \alpha_{A_2} A_{2i} + \alpha_{intx} \times A_{1i} \times A_{2i} \times U_i$ and the contrast of interest compares regimes $g = (A_1 = a_1, A_2 = a_2)$ and $g' = (A_1 = a_1', A_2 = a_2')$, 
\begin{align*}
\tilde{\phi}_i^{(b)} = 
&
 \alpha_{A_1}^{(b)} a_{1} + \alpha_U^{(b)} \tilde{u}^{g (b)}_i + \alpha_{A_2}^{(b)} a_2 + \alpha_{intx}^{(b)} a_1 \times a_2 \times \tilde{u}^{g (b)}_i 
 - \\
 & \alpha_{A_1}^{(b)} a_{1}' + \alpha_U^{(b)} \tilde{u}^{g' (b)}_i + \alpha_{A_2}^{(b)} a_2' + \alpha_{intx}^{(b)} a_1' \times a_2' \times \tilde{u}^{g' (b)}_i 
\end{align*}
\end{enumerate}
\item Calculate population estimate $ACE^{(b)} = \sum_{i=1}^{n_1} \tilde{\phi}_i^{(b)} / n_1$.
\item Construct a point estimate for $ACE$ as the posterior mean $\widehat{ACE}=\sum_{b=1}^{B} ACE^{(b)}/B$, and create quantile-based 95\% credible intervals as the $2.5^{th}$ and $97.5^{th}$ quantiles of $(ACE^{(1)}, \dots, ACE^{(B)})$.
\end{enumerate}

\subsection{Natural direct effects}
Suppose the true causal diagram underlying the mediation is shown in Figure~\ref{fig:meddag}. The unmeasured confounder $U$ can confound the (1) exposure-mediator, (2), exposure-outcome, or (3) mediator-outcome relationships, as well as any combination of (1) - (3).
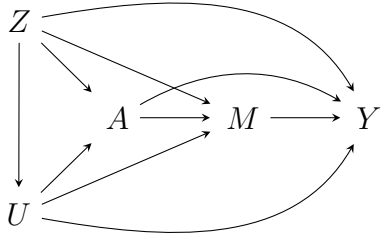
\begin{figure}
		\centering 
		\begin{tikzpicture}[ ->,shorten >=2pt,>=stealth,node distance=1cm,pil/.style={->,thick,shorten =2pt}]
			\node (a) {$A$};
            \node[above left=of a] (z) {$Z$};
            \node[below left=of a] (u) {$U$};
			\node[right=of a] (m) {$M$};
			\node[right=of m] (y) {$Y$};
            \draw[->] (z) to (a);
            \draw[->] (z) to (u);
            \draw[->] (u) to (a);
            \draw[->] (z) to (m);
            \draw[->] (u) to (m);
            \draw[->] (u) to [out=350, in=240] (y);
            \draw[->] (z) to [out=10, in=120] (y);
			\draw[->] (a) to [out=30, in=150] (y);
			\draw[->] (a) to (m);
			\draw[->] (m) to (y);
		\end{tikzpicture}
		\caption{Mediation causal structure with outcome $Y$, exposures $A$, mediator $M$, baseline confounder(s) $Z$, and unmeasured confounder $U$}
		\label{fig:meddagbaselineu}
\end{figure}

The population average natural direct effect of changing exposure $A$ to $a$ instead of $a^*$, while holding the mediator $M$ to its natural value under $A = a^*$, is given by $NDE = \E{Y^{a M^{a^*}} - Y^{a^* M^{a^*}}}$. This estimand has an intervention set $V = \{ A, M \}$ and can be formulated as a contrast in the regimes $g = (A = a, M = M^{a^*})$ and $g' = (A = a^*, M = M^{a^*})$.

\begin{enumerate}\label{alg:bdfnde}
\item Fit maximum likelihood models in the external data to obtain the prior $\pi(\theta)$ as detailed in Section~\ref{sec:bdfprior}.
\item Use NUTS with target probability distribution proportional to $\mathcal{L}_m \times \pi(\theta)$ in order to obtain posterior samples of the regression parameter vector $\theta$. For some large $B$, let  $\theta^{(1)}, \dots, \theta^{(B)}$ denote the $B$ posterior samples remaining after discarding warmup iterations.
\item \label{algstep:gsim} For MCMC iteration $b=1,\dots,B$ and $i = 1,\dots, n_1$:
\begin{enumerate}[a)]
\item Sample baseline covariate vector $\tilde{z}^{(b)}_i$ from the empirical distribution.
\item Sample $\tilde{u}^{(b)}_i$ in accordance with $\theta_U^{(b)}$ and $\tilde{z}_i^{(b)}$ using
\[ h_U^{-1} \left( \eta_U\left( \tilde{z}_i^{(b)}, \theta_U^{(b)}\right)\right) \]
In contrast to previous algorithms, this sampling does not need to be done for each $g_0 \in \{ g, g' \}$ because $U$ \emph{cannot} be a descendant of $A$ or $M$ in the causal graph for the natural direct effect to be well defined.
\item \label{algstep:medsamp} Sample mediator $\tilde{m}^{a^* (b)}_{i}$ according to $\theta_M^{(b)}$, $\tilde{z}_i^{(b)}$, and $\tilde{u}^{(b)}_i$ using
\[ h_M^{-1} \left( \eta_M\left( \tilde{z}_i^{(b)}, \tilde{u}^{ (b)}_i, a^*, \theta_M^{(b)}\right)\right) \]
\item For $g_0 \in \{g, g' \}$, draw $\tilde{y}^{g_0 (b)}_i$ in accordance with $g_0$, $\theta_Y^{(b)}$, $\tilde{z}^{(b)}_i$, $\tilde{u}^{(b)}_i$, and $\tilde{m}^{a^* (b)}_{i}$. For regime $g = (A = a, M = M^{a^*})$ this involves 
\[ h_Y^{-1} \left( \eta_Y\left( \tilde{z}_i^{(b)}, \tilde{u}^{ (b)}_i, \tilde{m}^{a^* (b)}_{i}, a, \theta_Y^{(b)}\right)\right) \]
while for $g' = (A = a^*, M = M^{a^*})$ the relevant equation will involve 
\[ h_Y^{-1} \left( \eta_Y\left( \tilde{z}_i^{(b)}, \tilde{u}^{ (b)}_i, \tilde{m}^{a^* (b)}_{i}, a^*, \theta_Y^{(b)}\right)\right) \]
Then calculate the individual-level causal effect
\begin{align*}
\tilde{\phi}_i^{(b)} = & \tilde{y}^{g (b)}_i - \tilde{y}^{g' (b)}_i
\end{align*}
Alternatively, if the conditional mean of $Y$ has a closed form $\mu(\theta_Y, z, u, a, m)$, define the individual-level causal effect as
\begin{align*}
\tilde{\phi}_i^{(b)} = & 
\mu\left(\theta_Y^{(b)}, \tilde{z}^{(b)}_i, \tilde{u}^{(b)}_i, a, \tilde{m}^{(b)}_i \right) - 
\mu\left(\theta_Y^{(b)}, \tilde{z}^{(b)}_i, \tilde{u}^{(b)}_i, a^*, \tilde{m}^{a^* (b)}_i \right).
\end{align*}
\end{enumerate}
\item Calculate population estimate $NDE^{(b)} = \sum_{i=1}^{n_1} \tilde{\phi}_i^{(b)} / n_1$.
\item Construct a point estimate for $NDE$ as the posterior mean $\widehat{NDE}=\sum_{b=1}^{B} NDE^{(b)}/B$, and create quantile-based 95\% credible intervals as the $2.5^{th}$ and $97.5^{th}$ quantiles of $(NDE^{(1)}, \dots, NDE^{(B)})$.
\end{enumerate}

\subsection{Natural indirect effects}\label{alg:bdfnie}
Again suppose that the correct causal diagram is as in Figure~\ref{fig:meddagbaselineu}, where natural direct effects are well defined.

The population average natural indirect effect is the effect of changing the mediator $M$ from the value it naturally takes under exposure $A = a^*$ to the value it naturally takes under $A=a$, while holding the exposure constant at level $a$. In potential outcome notation, this quantity is given by $NIE = \E{Y^{a M^{a}} - Y^{a M^{a^*}}}$. This estimand has an intervention set $V = \{ A, M \}$ and can be formulated as a contrast in the regimes $g = (A = a, M = M^{a})$ and $g' = (A = a, M = M^{a^*})$.

The estimation algorithm is the same as in Section~\ref{alg:bdfnde} until Step~\ref{algstep:gsim}, where it continues as follows.

\begin{enumerate}\setcounter{enumi}{-1}\addtocounter{enumi}{3}
\item For MCMC iteration $b=1,\dots,B$ and $i = 1,\dots, n_1$:
\begin{enumerate}[a)]
\item Sample baseline covariate vector $\tilde{z}^{(b)}_i$ from the empirical distribution.
\item Sample $\tilde{u}^{(b)}_i$ in accordance with $\theta_U^{(b)}$ and $\tilde{z}_i^{(b)}$ using
\[ h_U^{-1} \left( \eta_U\left( \tilde{z}_i^{(b)}, \theta_U^{(b)}\right)\right) \]
\item For each $a_0 \in \{a, a^*\}$, sample mediator $\tilde{m}^{a_0, (b)}_{i}$ according to $\theta_M^{(b)}$, $\tilde{z}_i^{(b)}$, and $\tilde{u}^{(b)}_i$ using
\[ h_M^{-1} \left( \eta_M\left( \tilde{z}_i^{(b)}, \tilde{u}^{ (b)}_i, a_0, \theta_M^{(b)}\right)\right) \]
\item For $g_0 \in \{g, g' \}$, draw $\tilde{y}^{g_0 (b)}_i$ in accordance with $g_0$, $\theta_Y^{(b)}$, $\tilde{z}^{(b)}_i$, $\tilde{u}^{(b)}_i$, and $\tilde{m}^{a_0 (b)}_{i}$ for the $a_0$ corresponding to the $M$ counterfactual in $g_0$. For regime $g = (A = a, M = M^{a})$ this involves 
\[ h_Y^{-1} \left( \eta_Y\left( \tilde{z}_i^{(b)}, \tilde{u}^{ (b)}_i, \tilde{m}^{a (b)}_{i}, a, \theta_Y^{(b)}\right)\right) \]
while for $g' = (A = a, M = M^{a^*})$ the relevant equation will involve 
\[ h_Y^{-1} \left( \eta_Y\left( \tilde{z}_i^{(b)}, \tilde{u}^{ (b)}_i, \tilde{m}^{a^* (b)}_{i}, a, \theta_Y^{(b)}\right)\right) \]
Then calculate the individual-level causal effect
\begin{align*}
\tilde{\phi}_i^{(b)} = & \tilde{y}^{g (b)}_i - \tilde{y}^{g' (b)}_i
\end{align*}
Alternatively, if the conditional mean of $Y$ has a closed form $\mu(\theta_Y, z, u, a, m)$, define the individual-level causal effect as
\begin{align*}
\tilde{\phi}_i^{(b)} = & 
\mu\left(\theta_Y^{(b)}, \tilde{z}^{(b)}_i, \tilde{u}^{(b)}_i, a, \tilde{m}^{a (b)}_i \right) - 
\mu\left(\theta_Y^{(b)}, \tilde{z}^{(b)}_i, \tilde{u}^{(b)}_i, a, \tilde{m}^{a^* (b)}_i \right).
\end{align*}
\end{enumerate}
\item Calculate population estimate $NIE^{(b)} = \sum_{i=1}^{n_1} \tilde{\phi}_i^{(b)} / n_1$.
\item Construct a point estimate for $NIE$ as the posterior mean $\widehat{NIE}=\sum_{b=1}^{B} NIE^{(b)}/B$, and create quantile-based 95\% credible intervals as the $2.5^{th}$ and $97.5^{th}$ quantiles of $(NIE^{(1)}, \dots, NIE^{(B)})$.
\end{enumerate}

\subsection{Randomized interventional analogs to the natural indirect effect}\label{alg:bdfrnie}
The estimation algorithm is the same as in Section~\ref{alg:bdfnde} until Step \ref{algstep:gsim}, where it continues as follows.
\begin{enumerate}\setcounter{enumi}{-1}\addtocounter{enumi}{3}
\item For MCMC iteration $b=1,\dots,B$ and $i = 1,\dots, n_1$:
\begin{enumerate}[a)]
\item Sample baseline covariate vector $\tilde{z}_i$ from the empirical distribution.
\item For each $g_0 \in \{ g, g' \}$ and $a_0 \in \{ a, a^*\}$, sample $\tilde{u}^{a_0, g_0 (b)}_i$ in accordance with $\theta_U^{(b)}$ and $\tilde{z}_i^{(b)}$ using
\[ h_U^{-1} \left( \eta_U\left( \tilde{z}_i^{(b)}, a_0, \theta_U^{(b)}\right)\right) \]
\item For $g_0 \in \{ g, g' \}$, sample randomized mediator $\tilde{m}^{g_0 (b)}_{i}$ in accordance with $\theta_U^{(b)}$, $\tilde{z}_i^{(b)}$, and $\tilde{u}_i^{a_0, g_0(b)}$. For regime $g = (A = a, M = H_z(a = a))$, draw $\tilde{m}^{g (b)}_{i}$ using
\[ h_M^{-1} \left( \eta_M\left( \tilde{z}_i^{(b)}, a, \tilde{u}_i^{a, g (b)},  \theta_M^{(b)}\right)\right) \]
and for $g' = (A = a, M = H_z(a = a^*))$ draw $\tilde{m}^{g' (b)}_{i}$ using
\[ h_M^{-1} \left( \eta_M\left( \tilde{z}_i^{(b)}, a^*, \tilde{u}_i^{a^*, g' (b)},  \theta_M^{(b)}\right)\right) \]
\item Define individual-level causal contrast For $g_0 \in \{g, g' \}$, draw $\tilde{y}^{g_0 (b)}_i$ in accordance with $g_0$, $\theta_Y^{(b)}$, $\tilde{z}^{(b)}_i$, $\tilde{u}^{(b)}_i$, and $\tilde{m}^{g_0 (b)}_{i}$. For regime $g = (A = a, M = H_z(a = a))$, draw $\tilde{y}^{g (b)}_{i}$ using
\[ h_Y^{-1} \left( \eta_Y\left( \tilde{z}_i^{(b)}, \tilde{u}^{a, g (b)}_i, \tilde{m}^{g (b)}_{i}, a, \theta_Y^{(b)}\right)\right) \]
while for $g' = (A = a, M = H_z(a = a^*))$, draw $\tilde{y}^{g' (b)}_{i}$ using
\[ h_Y^{-1} \left( \eta_Y\left( \tilde{z}_i^{(b)}, \tilde{u}^{a, g' (b)}_i, \tilde{m}^{g' (b)}_{i}, a, \theta_Y^{(b)}\right)\right) \]
Then calculate the individual-level causal effect
\begin{align*}
\tilde{\phi}_i^{(b)} = & \tilde{y}^{g (b)}_i - \tilde{y}^{g' (b)}_i
\end{align*}
Alternatively, if the conditional mean of $Y$ has a closed form $\mu(\theta_Y, z, u, a, m)$, define the individual-level causal effect as
\begin{align*}
\tilde{\phi}_i^{(b)} = & 
\mu\left(\theta_Y^{(b)}, \tilde{z}^{(b)}_i, \tilde{u}^{a, g (b)}_i, a, \tilde{m}^{g (b)}_i \right) - 
\mu\left(\theta_Y^{(b)}, \tilde{z}^{(b)}_i, \tilde{u}^{a, g' (b)}_i, a, \tilde{m}^{g' (b)}_i \right).
\end{align*}
\end{enumerate}
\item Calculate population estimate $rNIE^{(b)} = \sum_{i=1}^{n_1} \tilde{\phi}_i^{(b)} / n_1$.
\item Construct a point estimate for $rNIE$ as the posterior mean $\widehat{rNIE}=\sum_{b=1}^{B} rNIE^{(b)}/B$, and create quantile-based 95\% credible intervals as the $2.5^{th}$ and $97.5^{th}$ quantiles of $(rNIE^{(1)}, \dots, rNIE^{(B)})$.
\end{enumerate}

\subsection{Controlled direct effects}\label{alg:bdfcde}
The controlled direct effect is the effect of changing exposure $A$ to level $a$ from $a^*$ while holding the mediator $M$ fixed at level $m$, i.e., $CDE = \E{Y^{am} - Y^{a^*m}}$. In general, this requires no unmeasured exposure–outcome confounding and  no unmeasured mediator–outcome confounding. Both of these cases can be addressed by with slight modifications to previously stated versions of the BDF-SIM algorithm in order to obtain $B$ posterior samples $CDE^{(1)}, \dots, CDE^{(B)}$.

\subsubsection*{Exposure-outcome confounding or mediator-outcome confounding that not affected by treatment}
Suppose the true causal diagram is as in Figure~\ref{fig:meddagbaselineu}, where $U$ acts as an exposure-outcome confounder, mediator-outcome confounder, or both. (It may also be an exposure-mediator confounder, but the controlled direct effect is already identified if both the $U \to A$ and $U \to Y$ arrows are missing.)

For this causal structure, the controlled direct effect can be estimated using the algorithm from Section~\ref{alg:bdfnde}, replacing the stochastic assignment of $\tilde{m}^{a^* (b)}_i$ in Step~\ref{algstep:medsamp} with universal assignment to $m$ for all $i$ and $b$. 

\subsubsection*{Exposure-induced mediator-outcome confounding}
Suppose the true causal diagram is as in Figure~\ref{fig:bdfmed} in the main text, i.e., where $U$ is an exposure-induced mediator-outcome confounder. Further suppose scientific interest lies in the controlled direct effect of changing $A$ to level $a$ from $a^*$ while holding the mediator $M$ fixed at level $m$, i.e., $CDE = \E{Y^{am} - Y^{a^*m}}$.

The controlled direct effect can be estimated using the algorithm from Section~\ref{alg:bdftvc}, replacing $A_1$ with $A$ and $A_2$ with $M$. The two regimes are $g = (A = a, M = m)$ and $g' = (A = a^*, M = m)$.

\section{Data generation procedure for simulations}
The ``no-interaction'' case corresponding to no statistical interaction has $\Delta_{Y,AM}$ = 0.

When $U$ is not exposure-induced, $\Delta_{U,A}$ = 0.

For violations of transportability, $\beta_U = \alpha_U = 0$ was used for generation of the small data set. Otherwise, $\beta_U = \alpha_U = 1.5$ in order to induce strong mediator-outcome confounding by $U$.
\begin{align*}
Z_1 & \sim \mathrm{Bernoulli}(0.5) \\
Z_2 | Z_1 & \sim \mathrm{Bernoulli}(0.5) \\
A | Z_1, Z_2 & \sim \mathrm{Bernoulli}
\left( \logitinv{-0.2 + 0.5 Z_1 + 0.7 Z_2} \right)\\
U | A, Z_1, Z_2 & 
\sim \mathrm{Bernoulli}
\left( \logitinv{-0.4 + \Delta_{U,A} 1.5 A} \right)\\
M | U, A, Z_1, Z_2 & 
\sim \mathrm{Bernoulli}
\left( \logitinv{-1.5 + 0.3 Z_1 + 0.2 Z_2 + 0.7 A + \beta_U U} \right)\\
Y | M, U, A, Z_1, Z_2 & 
\sim \mathrm{Bernoulli}
\left( \logitinv{-2 + 0.3 Z_1 + 0.2 Z_2 + A + 0.8 M + \Delta_{Y, AM} A M + \alpha_U U} \right)
\end{align*}

\section{Frequentist bias corrections}
\subsection{$\delta$-$\gamma$ correction}
For each level of $z$, the bias due to $U$ for the $rNDE$ on the difference scale, assuming $A$ does not cause $U$ is
\begin{align}
B^{CDE}_{dg,add}(m=0|z) = & \delta_{m=0}(z) \gamma_{m=0}(z) \text{ with} \\
\delta_{m=0}(z) = & P(U = 1 | z, m=0, a = 1) - P(U = 1 | z, m=0, a = 0) \nonumber \\
\gamma_{m=0}(z) = & \E{Y=1| z, a, m=0, u = 1} - \E{Y=1| z, a, m=0, u = 1} \nonumber
\end{align}
The DG-corrected population estimate of the $rNDE$ is then given by
\begin{equation}
\widehat{rNDE}_{dg} = \sum_{z} \left( \widehat{rNDE}_{uc}(z) - B^{CDE}_{dg,add}(m=0|z) \right) p(z).
\end{equation}

\subsection{Interaction correction}
The bias in the additive $NDE$ from by a mediator-outcome confounder $U$ which is \emph{not} exposure-induced is given by:
\begin{align*}
B^{NDE}_{ix,add}(z) 
= &
\sum_{m,u} \bigg( 
\bigg[ \E{Y|a,m,z,u}P(u|a,m,z) - \E{Y|a^*,m,z,u}P(u|a^*,m,z) \bigg] P(m|a^*, c)
\bigg) \\
& - \sum_{m,u} \bigg(
\bigg[ \E{Y|a,m,z,u} - \E{Y|a^*,m,z,u}\bigg]P(m|a^*,z,u)P(u|z)
\bigg) 
\end{align*}
In our context, $a = 1$ and $a^* = 0$. To eliminate the possibility of model misspecification in the estimation of the bias correction term $B^{NDE}_{ix,add}(z)$, saturated parametric logistic regression models were adopted for $U$ and $M$ when possible. (For the data application, sparseness in the covariates made this impossible, and parametric models with many interaction terms were fit to reduce, but not eliminate, misspecification.) The model for $Y$ used to obtain $\E{Y|a=1,m,z,u}$ was the same logistic model used to obtain the naive $rNDE$, except with $U$ as an additional term.

The corrected estimate of the population $rNDE$ was calculated by
\begin{equation}
\widehat{rNDE}_{ix} = \sum_{z} \left( \widehat{rNDE}_{uc}(z) - B^{NDE}_{ix,add}(z) \right) p(z).
\end{equation}

\section{Additional simulation results}
\subsection{Credible interval widths}

Mean confidence and credible interval widths for the simulated scenarios are given in Table~\ref{tab:width}. The Bayesian estimators tend to have widths comparable to frequentist analogs when there is no transportability, but tends to have wider intervals when there is substantial bias in the external data. BDF-CF may perform slightly better than BDF-SIM (i.e., have narrower intervals) in small samples due to smaller Monte Carlo error.

\begin{table}[ht!]
\caption{\label{tab:width}Widths of 95\% confidence and credible intervals for
             naive, delta-gamma (DG) and interaction (IX) frequentist corrections, 
             simulation-based (BDF-SIM) and closed-form (BDF-CF) Bayesian data fusion estimators, 
             calculated in 200 replicates with exposure-induced mediator-outcome confounding}
\centering
\begin{tabular}[t]{lllrrrrr}
\toprule
Transportability & Interaction & Sample sizes & Naive & DG & IX & BDF-SIM & BDF-CF\\
\midrule
Yes & No & (150, 1500) & 0.101 & 0.100 & 0.100 & 0.108 & 0.107\\
Yes & No & (500, 5000) & 0.055 & 0.055 & 0.055 & 0.059 & 0.058\\
Yes & No & (1000, 10000) & 0.039 & 0.039 & 0.039 & 0.042 & 0.042\\
Yes & Yes & (150, 1500) & 0.101 & 0.098 & 0.101 & 0.112 & 0.110\\
Yes & Yes & (500, 5000) & 0.056 & 0.054 & 0.056 & 0.061 & 0.061\\
Yes & Yes & (1000, 10000) & 0.039 & 0.038 & 0.039 & 0.043 & 0.043\\
\addlinespace
No & No & (150, 1500) & 0.100 & 0.101 & 0.101 & 0.100 & 0.099\\
No & No & (500, 5000) & 0.055 & 0.055 & 0.055 & 0.054 & 0.054\\
No & No & (1000, 10000) & 0.039 & 0.039 & 0.039 & 0.038 & 0.038\\
No & Yes & (150, 1500) & 0.101 & 0.099 & 0.102 & 0.103 & 0.101\\
No & Yes & (500, 5000) & 0.056 & 0.054 & 0.055 & 0.056 & 0.054\\
No & Yes & (1000, 10000) & 0.039 & 0.038 & 0.039 & 0.039 & 0.038\\
\bottomrule
\end{tabular}
\end{table}

\subsection{Performance when unmeasured confounder is not exposure-induced}

\begin{figure}[ht!]
    \centering
    \includegraphics[width = 0.95\textwidth]{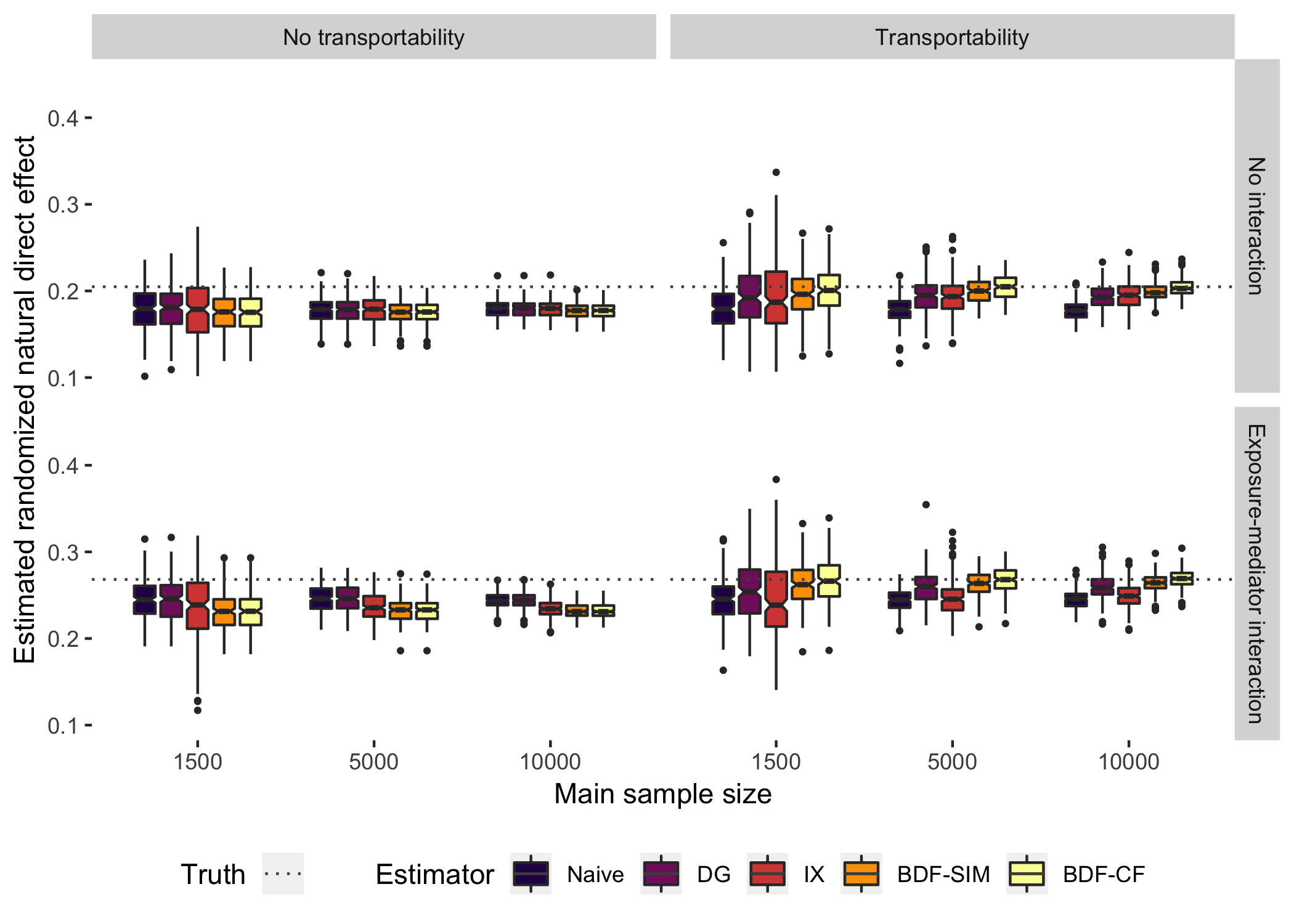}
    \caption{Randomized natural direct effects estimated with naive, delta-gamma (DG) correction, interaction (IX) correction, simulation-based Bayesian data fusion (BDF-SIM), and closed-form Bayesian data fusion (BDF-CF) estimators, with and without exposure-mediator interaction and causal transportability between main and external data sets.}
    \label{fig:simboxplotnouei}
\end{figure}

In the absence of transportability, all of the bias corrections considered perform poorly. However, the BDF-SIM and BDF-CF estimates are more confident about their (incorrect). In the top right panel of Figure~\ref{fig:simboxplotnouei}, one can see that the BDF approaches reduce bias better than the frequentist corrections in smaller sample sizes, but that the difference in performance is virtually eliminated with a main data set size of $n=10,000$. Interestingly, the IX correction does \emph{not} exhibit good performance when there is a strong exposure-mediator interaction, particularly in small samples.

\label{lastpage}

\end{document}